\documentstyle[eqsecnum, preprint, aps, fixes, epsfig]{revtex}
\tightenlines
\begin{document}

\def\citeNUM#1{b@#1}
\def\citenum#1{\expandafter\csname\citeNUM{#1}\endcsname}
\def\half{{\textstyle {1\over2}}}

\newsavebox{\lsim}
\savebox{\lsim}[1.0em]{\raisebox{-0.3ex}{$\stackrel{<}{\scriptstyle\sim}$}}

\preprint{UW/PT 99-06}
\title{From Instantons to Sphalerons: Time-dependent\\ Periodic
Solutions of $\mathbf{SU(2)}$-Higgs Theory}
\author{Keith L. Frost and Laurence G. Yaffe}
\address{University of Washington, Department of Physics,
Seattle, Washington 98105-1560}
\date{\today} 
\maketitle

\begin{abstract}
  We solve numerically for periodic, spherically symmetric, classical
  solutions of $\mathrm{SU}(2)$-Higgs theory in four-dimensional Euclidean
  space.  In the limit of short periods the solutions approach tiny
  instanton--anti-instanton superpositions while, for longer periods,
  the solutions merge with the static sphaleron.  A previously
  predicted bifurcation point, where two branches of periodic
  solutions meet, appears for Higgs boson masses larger than $3.091 \,
  M_W$.
\end{abstract}

\section{Introduction}

In electroweak theory, baryon number non-conserving processes are a
consequence of topological transitions in which there is an order one
change in the Chern-Simons number of the $\mathrm{SU}(2)$~gauge field.
At zero temperature, such transitions are quantum tunneling events,
and the rate of these transitions is directly related to the classical
action of instanton solutions of $\mathrm{SU}(2)$~gauge
theory~\cite{tHooft_76}.  At
sufficiently high temperatures,%
\footnote {%
  But below the cross-over (or critical) temperature where ``broken''
  electroweak symmetry is restored.
}
the dominant mechanism for baryon number violation involves classical
thermally activated transitions over the potential energy barrier
separating inequivalent vacuum states.  The configuration
characterizing the top of the barrier is the static sphaleron solution
of $\mathrm{SU}(2)$-Higgs theory~\cite{Klinkhamer&Manton_84}; the energy of
this solution controls the thermally-activated transition
rate~\cite{KuzminRubakov&Shaposhnikov_85,Arnold&McLerran_88}.

When one lowers the temperature from the sphaleron dominated regime,
the topological transition rate is related to the action of time-dependent
periodic classical solutions of the Euclidean field equations with a
period~$\beta$ equal to the inverse temperature.%
\footnote
    {%
      In $\mathrm{SU}(2)$-Higgs theory, configurations resembling the zero-size
      limit of instantons and anti-instantons also play a role in
      determining the transition rate at sufficiently low
      temperatures\cite{tHooft_76,HabibMottola&Tinyakov_96,Frost&Yaffe_99}.
    }
For convenience, we will refer to such solutions as periodic ``bounces''.%
\footnote
    {%
      Coleman~\cite{Coleman_77} used the term ``bounce'' to refer to a
      time-dependent Euclidean classical solution beginning and ending
      at a (metastable) local minimum of the potential.  Our usage is
      a generalization.
}%
\begin{figure}
\begin{center}
\setlength {\unitlength} {1cm}
\begin{picture}(0,0)
  \put(3.0,-0.5){\footnotesize (a) $M_H < 3.091 \, M_W$}
\end{picture}
\epsfysize=5.0cm
\epsfxsize=8.0cm
\epsfbox{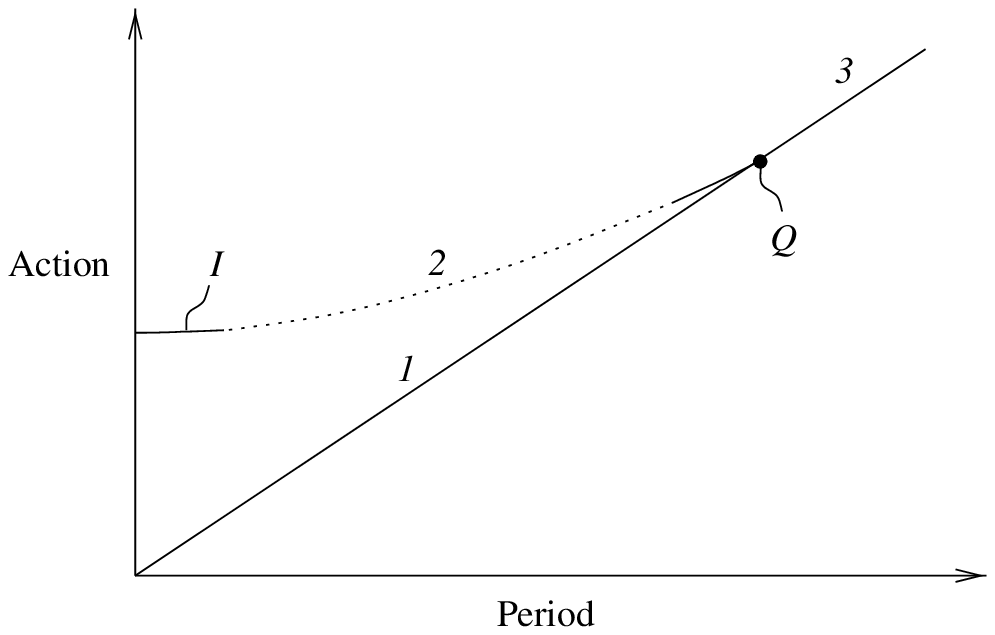}
\begin{picture}(0,0)
  \put(3.0,-0.5){\footnotesize (b) $M_H > 3.091 \, M_W$}
\end{picture}
\epsfysize=5.0cm
\epsfxsize=8.0cm
\epsfbox{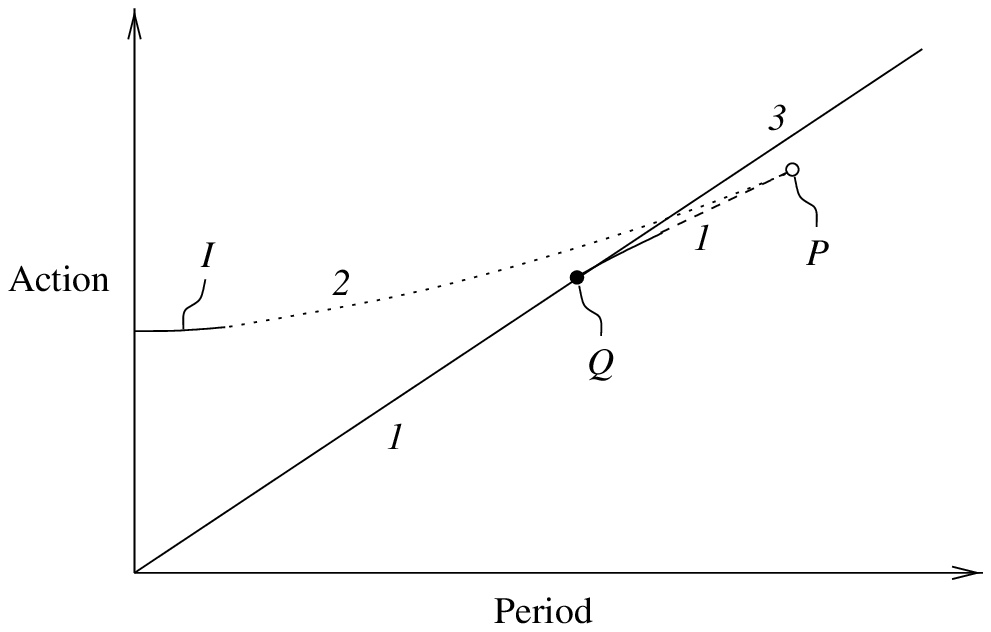}
\end{center}
\bigskip
\caption
{The simplest consistent scenario for the behavior of periodic
  solutions in $\mathrm{SU}(2)$-Higgs theory.  The straight line through the
  origin represents the static sphaleron, whose action is simply its
  energy times the period.  The line labeled $I$ denotes the branch of
  instanton--anti-instanton solutions whose action, in the limit of
  vanishing period, approaches $4\pi/\alpha_W$.  The number next to
  each curve indicates the number of unstable modes of the solution.
  Plot (a) illustrates the situation when the Higgs mass is less than
  $3.091 \, M_W$.  A branch of bounce solutions emerges from the
  sphaleron at a critical period, and moves toward shorter period and
  lower action as the deviation from the sphaleron increases.  As
  indicated by the dotted line, it is consistent to imagine that these
  limiting forms are the two ends of a single branch of solutions.
  Plot (b) illustrates the situation when the Higgs mass is larger
  than $3.091 \, M_W$.  In this case, a branch of bounce solutions
  emerges from the sphaleron at the critical period, and moves toward
  longer period and larger action as the deviation from the sphaleron
  increases.  The simplest consistent scenario is for a single
  additional bifurcation, marked $P$, to exist, at which the branch of
  bounces coming from the sphaleron merges with the branch of
  instanton--anti-instanton solutions.
\label{hypo-fig}
}
\end{figure}

In a previous paper, we used time-dependent classical perturbation
theory to study periodic bounce solutions which are small deformations
of the static sphaleron~\cite{Frost&Yaffe_99}.  We found that for
sufficiently small Higgs boson mass, $M_H < 3.091\,M_W$, bounce
solutions which are small oscillations about the sphaleron have an
action (per period) larger than that of the sphaleron.  In a plot of
action versus period, this branch of bounce solutions lies above the
static sphaleron line, and merges with the sphaleron at its critical
period, as shown in Fig.~\ref{hypo-fig}a.  For short periods, $\beta
\ll M_W^{-1}$, one can find the asymptotic form of periodic solutions
resembling an alternating array of instantons and anti-instantons,
where the attractive instanton--anti-instanton interactions exactly
counterbalance the tendency of each instanton and anti-instanton to
shrink~\cite{KhlebnikovRubakov&Tinyakov_91,Frost&Yaffe_99}.  In the
short-period limit, the resulting instanton size is
\begin {equation}
    \rho(\beta) = \sqrt{2} M_W \left( {\beta \over 2\pi} \right)^2
    \left[1+{\mathcal O}(M_W\beta)\right],
\label{rho}
\end {equation}
and the action (per period) is
\begin {equation}
    S(\beta) = {1 \over g^2}
\left[ 16\pi^2 + \frac{M_W^4 \beta^4}{2 \pi^2}
+ {\mathcal O}(M_W^5\beta^5) \right]\,.
\label{action-ii}
\end {equation}
Both the bounce solutions, and these instanton--anti-instanton
solutions, have two negative modes in their small fluctuation spectra
\cite{Frost&Yaffe_99}.  For $M_H < 3.091 \, M_W$, it is reasonable to
conjecture that a single branch of periodic solutions joins these two
limiting forms, as illustrated in Fig.~\ref{hypo-fig}a.

For Higgs boson masses above $3.091\,M_W$, the branch of bounce
solutions which emerge from the sphaleron extends to longer period
from the sphaleron bifurcation point, and the action of these bounces
lies below the action of the sphaleron (for the same period), as
shown in Fig.~\ref{hypo-fig}b.  These bounce solutions have a single
negative mode in their small fluctuation
spectra~\protect\cite{Frost&Yaffe_99}.  It is impossible for
short-period instanton--anti-instanton solutions to smoothly join with
these bounce solutions.  The simplest consistent scenario, illustrated
in Fig.~\ref{hypo-fig}b, is for there to be one additional bifurcation
point (in the space of spherically symmetric, periodic solutions) at
which the bounce solutions from the sphaleron merge with the
instanton--anti-instanton solutions.

In this paper, we present results of a numerical calculation of
periodic, spherically-symmetric, bounce solutions in
$\mathrm{SU}(2)$-Higgs theory.  Our goal was to find, and follow, the
different branches of periodic bounce solutions as both the period,
and the Higgs mass, are varied, and to see how these solutions
interpolate between the sphaleron and small instanton--anti-instanton
superpositions.  We found that the conjectured behavior illustrated in
Fig.~\ref{hypo-fig} is, in fact, correct.  While preparing this
manuscript, it came to our attention that similar work has been done
by Bonini~\emph{et al.}~\cite{Bonini&al_99}.  Their results are
consistent with ours.

We considered $\mathrm{SU}(2)$~gauge theory in $3{+}1$~dimensions with
a single Higgs scalar in the fundamental representation.  This model
represents the bosonic sector of the standard model of the weak
interactions in the limit of small weak mixing angle.  The action may
be written in the form%
\footnote{For convenience, we have rescaled the gauge and Higgs fields
  so that the action has an overall~$1/g^2$.  The conventional Higgs
  vacuum expectation value is~$v=2M_W/g$, and the usual quartic
  coupling is~$\lambda = \frac{1}{8}g^2 M_H^2 / M_W^2$.  As usual, the
  weak fine-structure constant is~$\alpha_W \equiv g^2/4\pi$.}
\begin{equation}
  S = \frac{1}{g^2} \int d^4 x \left[ -\frac{1}{2} \, \mbox{Tr}
    (F_{\mu\nu}F^{\mu\nu}) + (D_\mu\Phi)^\dagger D^\mu\Phi + \frac{M_H^2}{8
      M_W^2}(\Phi^\dagger\Phi - 2M_W^2)^2 \right].
\label{action-4d}
\end{equation}
The scalar field~$\Phi$ is an $\mathrm{SU}(2)$~doublet with covariant
derivative~$D_\mu = (\partial_\mu + A_\mu)$, where~$A_\mu \equiv A^a_\mu
\tau^a / 2 i$.  The gauge field strength is the commutator~$F_{\mu\nu}
= [D_\mu, D_\nu]$.
When restricted to spherically symmetric fields,
the action~(\ref{action-4d}) reduces to the two-dimensional
form~\cite{Ratra&Yaffe_88} 
\begin{eqnarray}
  S & = & \frac{4 \pi}{g^2} \int dr \, d\tau \, \left[ \frac{1}{4} \, r^2
    f_{\mu\nu}f^{\mu\nu} + |D \chi|^2 + r^2 |D \phi|^2 +
    \frac{M_H^2}{8 M_W^2} \, r^2(|\phi|^2{-}2 M_W^2)^2 \right. \nonumber \\ 
  & & \kern 1.6in \left. {}  + \frac{1}{2 r^2}(|\chi|^2{-}1)^2
    + \frac{1}{2} |\phi|^2(|\chi|^2{+}1) - {\mathrm Re}(\chi^*\phi^2) 
        \right]\,,
\label{action-2d}
\end{eqnarray}
where $\phi$ is a complex scalar which represents the spherically
symmetric Higgs field, $\chi$ is a complex scalar representing the
spherically symmetric transverse modes of the gauge fields, and
$f_{\mu\nu}$ is a two-dimensional $\mathrm{U}(1)$ field strength tensor.  After
reduction to spherically symmetric form, the original classical
$\mathrm{SU}(2)$ gauge theory becomes a classical $\mathrm{U}(1)$ gauge theory
in 1+1 dimensions.  The field $\phi$ has a $\mathrm{U}(1)$ charge of $1/2$,
while $\chi$ has $\mathrm{U}(1)$ charge $1$.  Note that finite action
configurations must satisfy~$|\chi|\to 1$ as~$r\to 0$, and~$|\phi|\to
\sqrt{2}M_W$,~$|\chi|\to 1$ with~$\chi^*\phi^2$ real and positive
as~$r\to\infty$.

\section{Numerical Methods}

To perform calculations with the two-dimensional action
(\ref{action-2d}), we chose to fix radial gauge, $a_1 = 0$.  We used a
spatial discretization which is uniform in the transformed variable
\begin{equation}
s = \ln \left[ \frac{1 + M r}{1 + m r} \right] \Big{/} \ln (M / m)\,,
\label{s-def}
\end{equation}
with
\begin{eqnarray}
m &\equiv& \max \left\{ \half \rho(\beta)^{-1},\,\half M_W \right\} ,
\nonumber \\
M &\equiv& \max \left\{ 2m,\, M_H \right\} .
\end{eqnarray}
The transformation (\ref {s-def}) maps the semi-infinite line $0 \le r
< \infty$ onto the unit interval $0 \leq s < 1$, with an approximately
logarithmic distribution of points in $r$ for scales in between
$M^{-1}$ and $m^{-1}$.  The long distance scale $m^{-1}$ was set to
the shorter of $2M_W^{-1}$ and $2\rho(\beta)$, where $\rho(\beta)$,
given in Eq.~(\ref {rho}), is the leading-order size of an instanton
stabilized against collapse in the instanton--anti-instanton periodic
solution with period~$\beta$.  The short distance scale $M^{-1}$ was
chosen to equal the shorter of $m^{-1}/2$ and $M_H^{-1}$.  We
found that this choice of discretization smoothly encompassed the
different spatial scales which are most relevant in periodic bounce
solutions as the parameters $M_H$ and $\beta$ are varied.  After
changing variables from $r$ to $s$, spatial derivatives in the action
(\ref {action-2d}) were replaced by nearest-neighbor finite
differences.  Calculations were performed using $L=64$ points for the
spatial discretization, and the resulting discretization error in, for
example, the sphaleron mass was less than one part in $10^4$.

Since we were looking for solutions with periodic time dependence, we
chose to expand the time dependence of fields in a (truncated) Fourier
series.  For bounce solutions which are close to the static sphaleron,
extremely accurate calculations can be performed using only
a few Fourier components in each field.%
\footnote
{%
  In classical perturbation theory about the sphaleron, the $n$-th
  harmonic of the fundamental frequency is only generated at $n$-th
  order in time-dependent perturbation theory\cite{Frost&Yaffe_99}.  
}    
As these solutions are followed toward very short periods, one finds
that they approach instanton--anti-instanton superpositions in which
the instanton (and anti-instanton) size shrinks quadratically with
period as shown in Eq.~(\ref{rho}).  Consequently, as the period
decreases, an increasing number of Fourier components are needed to
accurately represent the time dependence.

For numerical purposes, it is convenient to separate the
complex fields $\chi$ and $\phi$ into real and imaginary parts,
\begin {equation}
    \chi \equiv \alpha + i \beta \,, \qquad
    \phi \equiv \mu + i \nu \,.
\end {equation}
Table \ref{fourier-table} shows which fields change sign under the
action of the discrete symmetries of time-reversal and parity [which
is $\mathrm{U}(1)$ charge conjugation in the 1+1 dimensional theory
(\ref {action-2d})], and the resulting form of the truncated Fourier
series.  The bounce solutions are time reversal invariant, so only
cosine terms are needed for the time-reversal even fields, and only
sine terms for the time-reversal odd field $a_0$.  The solutions are
also invariant under the combination of parity plus time translation
by half a period.%
\footnote
{%
  These symmetries imply that the solutions reach ``turning-point''
  configurations, in which the time-reversal odd fields vanish, twice
  during each period.  The two turning-point configurations are parity
  reflections of each other, and occur half a period apart.  
}  
Because of this additional discrete symmetry,%
\footnote
{%
  In our previous paper~\cite{Frost&Yaffe_99}, we described
  the effect of discrete symmetries on the negative modes of the
  bounce and instanton--anti-instanton solutions incorrectly.  Both
  negative modes of the instanton--anti-instanton and the bounce are
  time-reversal even.  But the negative mode of the
  instanton--anti-instanton given by moving the instanton and
  anti-instanton closer together in imaginary time is odd under the
  combination of parity with time translation by half a period.  The
  static negative mode of the sphaleron (and hence the quasi-static
  negative mode of the bounce) is also odd under parity combined with
  $\beta/2$ time translation.  The other negative mode of the
  instanton--anti-instanton, given by a symmetric alteration of the
  sizes of the instanton and anti-instanton, is even under parity
  combined with $\beta/2$ time translation, as is the second
  negative mode (corresponding to changing the amplitude of
  oscillations about the sphaleron) of the branch of bounces with two
  negative modes.
}
only even frequencies appear in the Fourier expansions of parity even
fields, and only odd frequencies in the Fourier series for parity odd
fields.  Each Fourier series was truncated after $N$ (non-zero) terms;
practical computations ranged from $N=4$ to $N=64$ terms.

\begin{table}
\begin{center}
\setlength{\tabcolsep}{4.0\tabcolsep}
\renewcommand{\arraystretch}{1.7}
\begin{tabular}{|c|c|c|l|}
\hline
{\bf Field}$^{\strut}$ & {\bf P} & {\bf T} &
        {\bf Truncated Fourier Series} \\*[0.4ex]
\hline
$\alpha(r,\tau)$ & $+$ & $+$ & $\displaystyle \sum_{n=0}^{N{-}1} 
                             \alpha^{(2n)}(r)\,\cos[2 n \omega \tau]
                             ^{\vphantom{\bigg|}}$
                             \\*[2.0ex]
$\mu(r,\tau)$    & $+$ & $+$ & $\displaystyle \sum_{n=0}^{N{-}1} 
                              \mu^{(2n)}(r)\,\cos[2 n \omega \tau]$\\*[2.0ex]
$\beta(r,\tau)$  & $-$ & $+$ & $\displaystyle \sum_{n=0}^{N{-}1} 
                    \beta^{(2n{+}1)}(r) \,\cos[(2n{+}1)\omega \tau]$\\*[2.0ex]
$\nu(r,\tau)$    & $-$ & $+$ & $\displaystyle \sum_{n=0}^{N{-}1}
                      \nu^{(2n{+}1)}(r) \,\cos[(2n{+}1)\omega \tau]$\\*[2.0ex]
$a_0(r,\tau)$    & $-$ & $-$ & $\displaystyle \sum_{n=0}^{N{-}1}
                      a_0^{(2n{+}1)}(r) \,\sin[(2n{+}1)\omega \tau]$\\*[2.0ex]
\hline
\end{tabular}
\end{center}
\caption{
Symmetry properties, and truncated Fourier series, of two-dimensional
real field components.  
$P$~is parity (or equivalently $\mathrm{U}(1)$~charge conjugation), $T$~is time
reversal, and a $+$~or~$-$ indicates whether the field is even or odd,
respectively, under the symmetry.  The fundamental angular
frequency~$\omega\equiv 2\pi/\beta$, where~$\beta$ is the period.
Only even harmonics occur in the expansions of parity even fields,
and only odd harmonics appear in parity odd fields.
\label{fourier-table}
}
\end{table}

Once the radial dependence of the fields is replaced with an $L$-point
discretization, and the time dependence restricted to an $N$-term
Fourier series, the action (\ref{action-2d}) becomes a function of a
finite number ($5NL$) of variables.  The gradient of the action,
$\delta S$, is a $5NL$-component vector, and the curvature of the
action, $\delta^2 S$, is a $5NL$-dimensional matrix with block
band-diagonal structure.  An iterative Newton method, based on steps
for which
\begin {equation}
    \delta({\rm fields}) = - (\delta^2 S)^{-1} \cdot \delta S \,,
\label {newton}
\end {equation}
was used to find stationary points of the action.  The block
band-diagonal structure of the curvature matrix $\delta^2 S$ was used
to minimize computation and memory requirements for the calculation.

For stability of the calculation, especially near bifurcations, it was
important to guard against taking Newton steps that were too large.
(That is, larger than the domain of validity of a local quadratic
approximation to the action, upon which the Newton iteration (\ref
{newton}) is based.)  We adopted the strategy of scaling down the size
of a step as needed to guarantee that the magnitude of the gradient of
the action actually decreased upon taking each step.

For each value of the Higgs mass, to find the branch of solutions
which merge with the sphaleron, we first found the sphaleron solution
\cite{Klinkhamer&Manton_84,Yaffe_89,Kunz&Brihaye_89,Frost&Yaffe_99},
and set the period $\beta$ equal to the period of infinitesimal
oscillations about the sphaleron, as calculated by time-dependent
perturbation theory~\cite{Frost&Yaffe_99}.  We then shifted the period
$\beta$ a small amount, in the direction indicated by the perturbative
calculation, and added the first- and second-order perturbative terms
to the sphaleron fields to construct an initial guess for the
oscillating solution.  Newton's iteration steps were then taken until
the fields converged, and the gradient vanished, to within the limits
of numerical accuracy.  A small modification of the period $\beta$ was
then introduced, and Newton's method again used to solve for the
classical fields at the new period.  We used linear extrapolation in
$\beta$ to guess trial values of the fields after each change in
period.  The step size in $\beta$ was automatically adjusted to the
largest change which would allow Newton's method to converge in a
reasonable number of steps using linearly extrapolated starting
points.

As calculations ventured farther from the sphaleron solution,
additional Fourier components were required to accurately represent
the time-dependence of the fields.  A useful check on the accuracy of
the calculation proved to be the computation of the continuum
Euclidean conserved energy as a function of time.  For continuum
classical fields, it is of course conserved.  As the cutoff in the
Fourier modes becomes a more significant source of error, small
fluctuations in the energy grow.  We set a threshold of $\delta E / E
\sim 10^{-4}$ in the relative fluctuations in the energy which the
accuracy of our calculation allowed.  When the fluctuations in the
energy surpassed this, we doubled the number of Fourier components
used in the calculation, up to a maximum of $N=64$ components.

To find branches of solutions which were not directly connected to the
sphaleron, we first followed, for small Higgs mass $M_H \sim M_W$, the
branch of solutions which is connected to the sphaleron down to
periods $\beta \sim M_W^{-1}$, where they approach isolated instantons
and anti-instantons.  These solutions are relatively insensitive to
the Higgs boson mass.  It was then a simple matter to increase the
Higgs boson mass into the range $M_H > 3.091\,M_W$, and then follow
the solutions back toward longer period (where they no longer merge
directly with the sphaleron) using the same linear extrapolation and
Newton iteration techniques to vary the period $\beta$.

\section{Results}

\begin{figure}
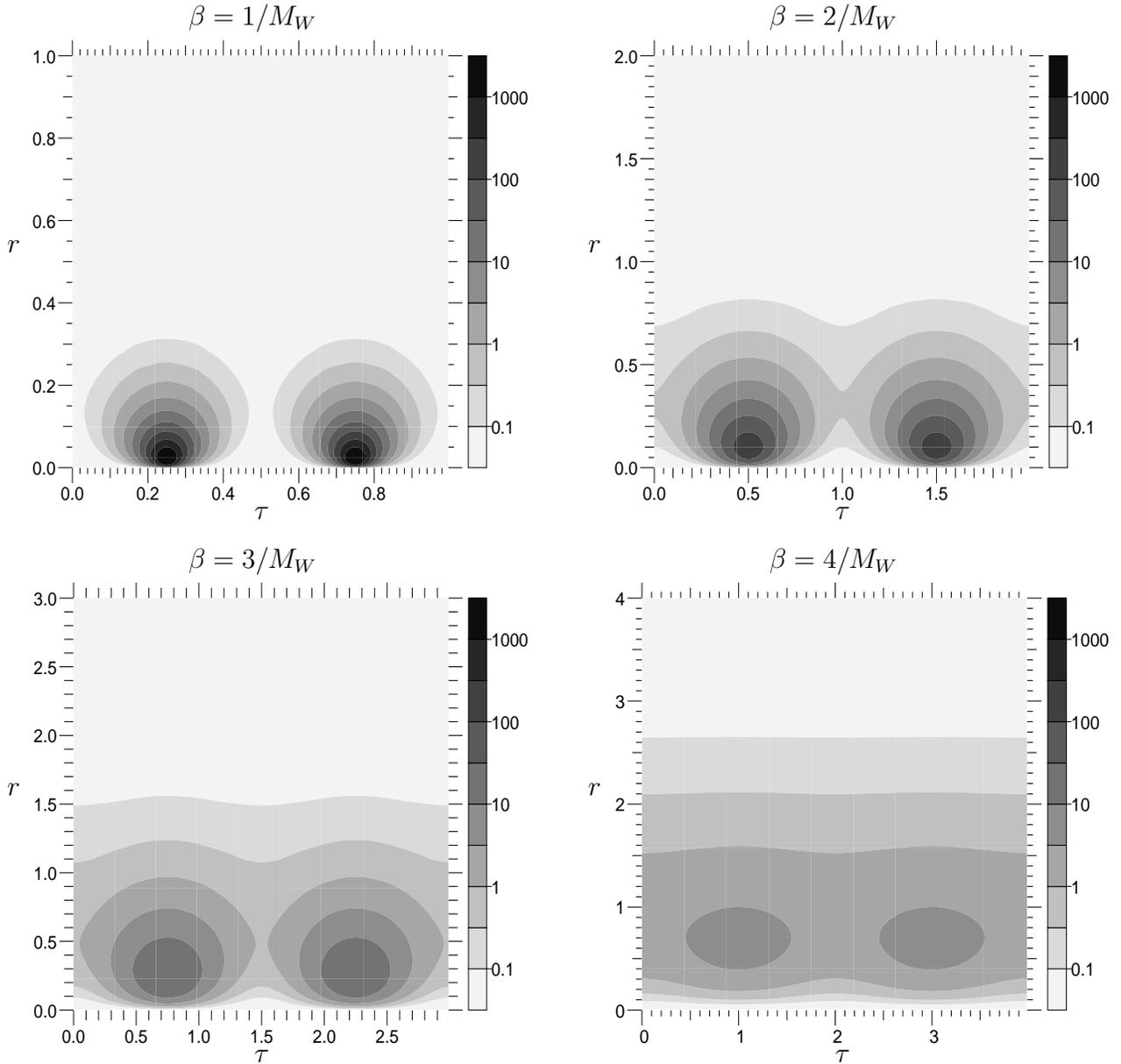

\vspace{0.3cm}
\begin{center}
\setlength{\unitlength}{1cm}
\begin{picture}(0,0)
  \put(-0.25, 3.75){$r$}
  \put(3.50, -0.3){$\tau$}
  \put(2.50, 7.2){$\beta=1/M_W$}
\end{picture}
\epsfysize=7.0cm
\epsfxsize=7.5cm
\epsfbox{act1.epsi}
\hfill
\begin{picture}(0,0)
  \put(-0.25, 3.75){$r$}
  \put(3.50, -0.3){$\tau$}
  \put(2.50, 7.2){$\beta=2/M_W$}
\end{picture}
\epsfysize=7.0cm
\epsfxsize=7.5cm
\epsfbox{act2.epsi}
\\[1.2cm]
\begin{picture}(0,0)
  \put(-0.25, 3.75){$r$}
  \put(3.50, -0.3){$\tau$}
  \put(2.50, 7.2){$\beta=3/M_W$}
\end{picture}
\epsfysize=7.0cm
\epsfxsize=7.5cm
\epsfbox{act3.epsi}
\hfill
\begin{picture}(0,0)
  \put(-0.25, 3.75){$r$}
  \put(3.50, -0.3){$\tau$}
  \put(2.50, 7.2){$\beta=4/M_W$}
\end{picture}
\epsfysize=7.0cm
\epsfxsize=7.5cm
\epsfbox{act4.epsi}
\end{center}
\vspace{0.5cm}
\caption[Action Density Contour Plots]{Contour plots of
  the two-dimensional action density of bounce solutions of
  $\mathrm{SU}(2)$-Higgs theory, as functions of radius $r$ and
  imaginary time $\tau$.  The Higgs boson mass is set to $M_H = M_W$,
  and the units are fixed to $M_W=1$.  The density shown is the
  integrand in the expression for the two-dimensional
  action~(\ref{action-2d}).  For Higgs boson mass $M_H < 3.091\,M_W$,
  shorter periods correspond to progressively more pronounced
  oscillations in imaginary time, culminating in tiny, well-separated
  instanton--anti-instanton configurations for $\beta \sim 1/M_W$, as
  shown at the upper left.  For Higgs boson mass $M_H = M_W$, at a
  period $\beta_0 \approx 4.1695 / M_W$, the periodic solutions merge
  with the static sphaleron.  The configuration at the lower right,
  with $\beta=4/M_W$, consists of relatively small oscillations about
  the static sphaleron.
\label{contour-fig}}
\end{figure}

We display in Fig.~\ref{contour-fig} representative contour plots
showing the action density of periodic solutions at four different
periods, with the Higgs boson mass $M_H=M_W$.  For Higgs boson masses
this small, as discussed above, the more pronounced the oscillation in
imaginary time is, the shorter the period becomes.
Figure~\ref{contour-fig} shows the gradual progression from small,
well-separated instanton--anti-instanton solutions with period $\beta
\sim 1/M_W$, up to fairly small oscillations about the static
sphaleron with period $\beta = 4/M_W$.  For $M_H=M_W$, the periodic
bounce solutions merge with the static sphaleron solutions at the
period $\beta_0 \approx 4.1695/M_W$.

\begin{figure}
\begin{center}
\setlength {\unitlength}{1cm}
\begin{picture}(0,0)
  \put(-1.0,3.5){$\alpha_W\,S$}
  \put(5.0,-0.5){$M_W\,\beta$}
  \put(1.5,5.0){$M_H = M_W$} 
\end{picture}
\epsfysize=6.2cm
\epsfxsize=10cm
\epsfbox{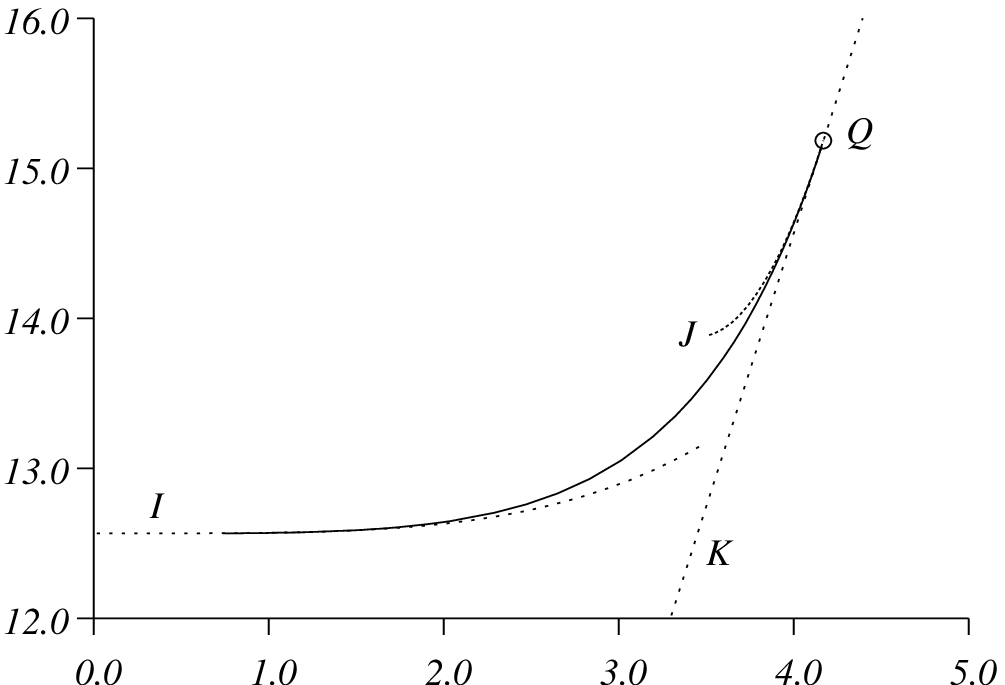}
\\[1.0cm]
\begin{picture}(0,0)
  \put(-1.0,3.5){$\alpha_W\,S$}
  \put(5.0,-0.5){$M_W\,\beta$}
  \put(1.5,5.0){$M_H = 3\,M_W$}
\end{picture}
\epsfysize=6.2cm
\epsfxsize=10cm
\epsfbox{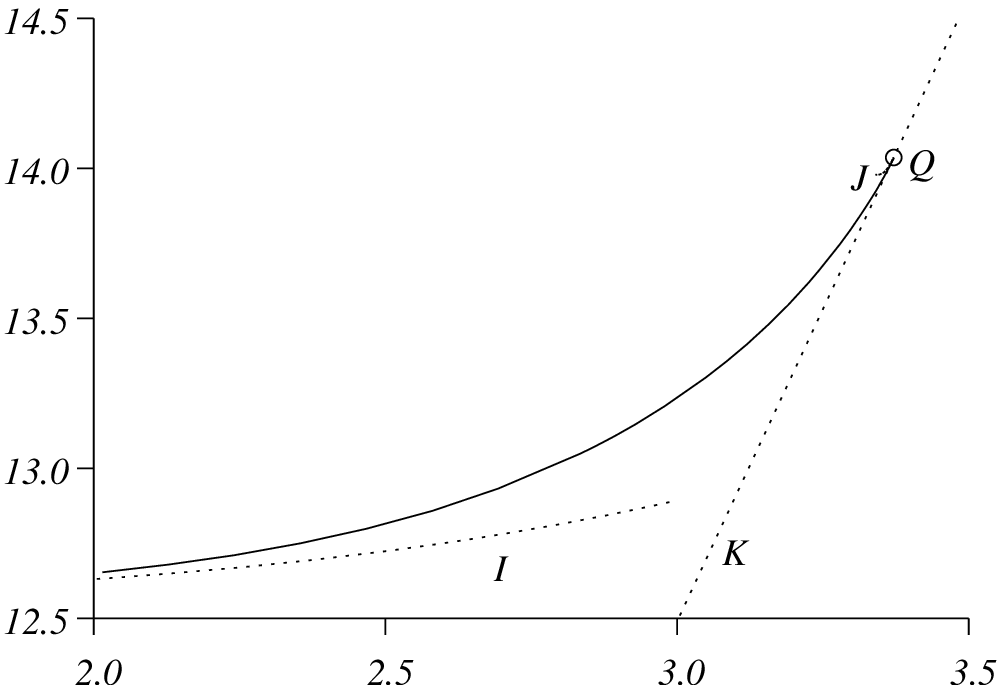}
\end{center}
\vspace{0.75cm}
\caption[$S$ vs $\beta$, $M_H$ small]{The action versus the period of
  $\mathrm{SU}(2)$-Higgs bounce solutions are plotted as solid curves.
  Dotted curves $I$ show the asymptotic forms for
  instanton--anti-instanton solutions.  Dotted curves $J$ indicate the
  leading-order calculation of small oscillations about the sphaleron,
  while dotted line $K$ is the sphaleron solution.  The small
  oscillations $J$ merge with the sphaleron $K$ at bifurcation point
  $Q$.  The upper graph shows a wide range of periods for $M_H = M_W$,
  while the lower graph provides a more detailed view near the
  bifurcation point $Q$ for $M_H = 3\,M_W$.
\label{svb1-fig}}
\end{figure}

Figure \ref{svb1-fig} contains two plots of action versus period,
which show the periodic Euclidean classical solutions of
$\mathrm{SU}(2)$-Higgs theory for $M_H=M_W$ and $M_H = 3\,M_W$,
respectively.  Note that the instanton--anti-instanton limit $I$ is
connected smoothly to the small oscillations $J$ about the sphaleron
$K$, by the solid curve of bounce solutions.  This simple picture
undergoes no fundamental change until $M_H$ exceeds $3.091\,M_W$.

\begin{figure}
\begin{center}
\setlength {\unitlength} {1cm}
\begin{picture}(0,0)
  \put(-1.0,3.5){$\alpha_W\,S$}
  \put(5.0,-0.5){$M_W\,\beta$}
  \put(1.5,5.0){$M_H = 4\,M_W$}
\end{picture}
\epsfysize=6.2cm
\epsfxsize=10.0cm
\epsfbox{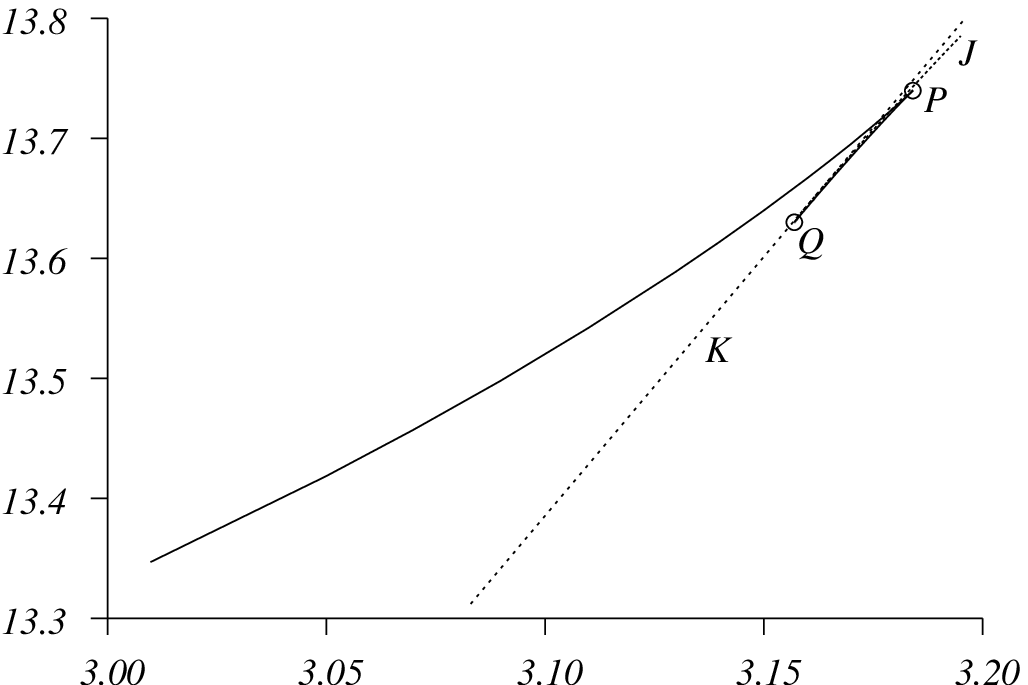}
\\[1.0cm]
\begin{picture}(0,0)
  \put(-1.0,3.5){$\alpha_W\,S$}
  \put(5.0,-0.5){$M_W\,\beta$}
  \put(1.5,5.0){$M_H = 6.665\,M_W$}
\end{picture}
\epsfysize=6.2cm
\epsfxsize=10.0cm
\epsfbox{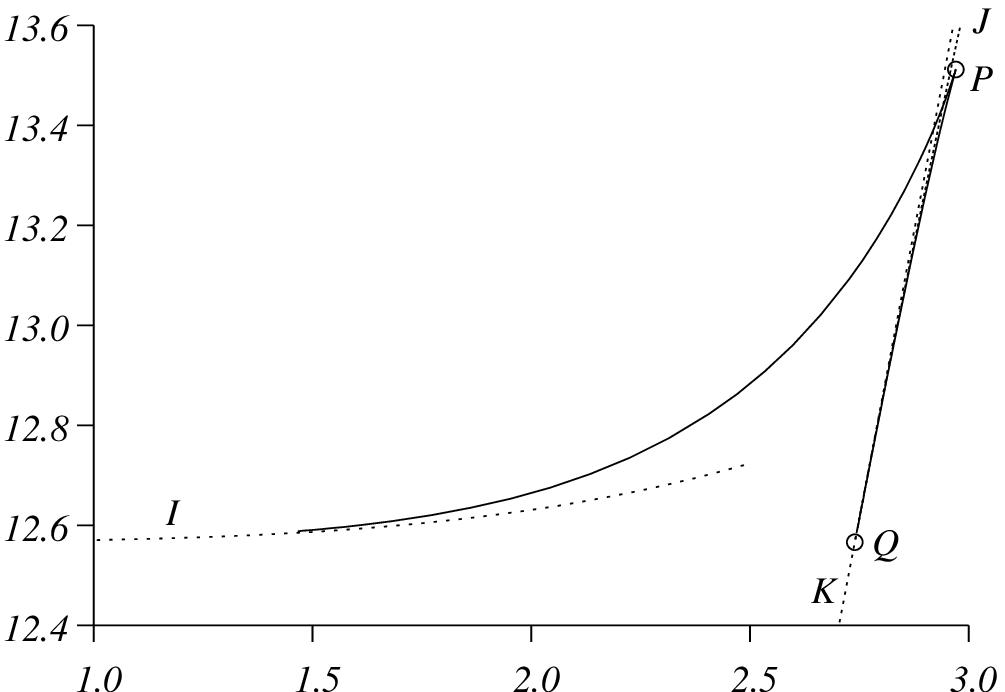}
\end{center}
\vspace{0.75cm}
\caption[$S$ vs $\beta$, $M_H=6.665\,M_W$]{Plots of action versus period of
  $\mathrm{SU}(2)$-Higgs bounce solutions for Higgs boson masses $M_H
  = 4\,M_W$ and $M_H=6.665\,M_W$. The solid curves, meeting at
  bifurcation point $P$, are the two branches of bounce solutions.
  Dotted curve $I$ shows the asymptotic form of the
  instanton--anti-instanton solutions.  Dotted curves $J$ indicate the
  leading-order calculation of small oscillations about the sphaleron,
  while dotted line $K$ is the sphaleron solution.  The small
  oscillations $J$ merge with the sphaleron $K$ at bifurcation point
  $Q$, whose action, at the Higgs boson mass $M_H \approx 6.665\, M_W$,
  is the same as that of the zero size limit of the
  instanton--anti-instantons $I$.  Bifurcation point $P$ was predicted
  to exist, by the perturbative calculation shown as $J$, for all
  Higgs boson masses $M_H > 3.091\,M_W$; we located it explicitly
  for a Higgs boson mass as small as $M_H = 3.2\,M_W$.  
\label{svb2-fig}}
\end{figure}

When the Higgs boson mass $M_H > 3.091\, M_W$, a second bifurcation
point $P$ appears, dividing the bounce solutions into two curves on a
plot of action vs.~period.  We calculated this bifurcation point
explicitly for $M_H=3.2\,M_W$, and for several larger Higgs boson
masses.  Figure~\ref{svb2-fig} shows the two branches of bounce
solutions as solid curves on plots of action versus period, for $M_H =
4\,M_W$ and $M_H = 6.665\,M_W$.  For $M_H > 6.665\,M_W$, the original
bifurcation $Q$, which joins the bounce solutions to the sphaleron,
has an action below the singular instanton--anti-instanton
configuration.
 
\begin{figure}
\begin{center}
\setlength {\unitlength} {1cm}
\begin{picture}(0,0)
  \put(-1.0,2.5){$M_W\,\beta$}
  \put(3.75,-0.5){$M_H / M_W$}
\end{picture}
\epsfysize=5cm
\epsfxsize=8cm
\epsfbox{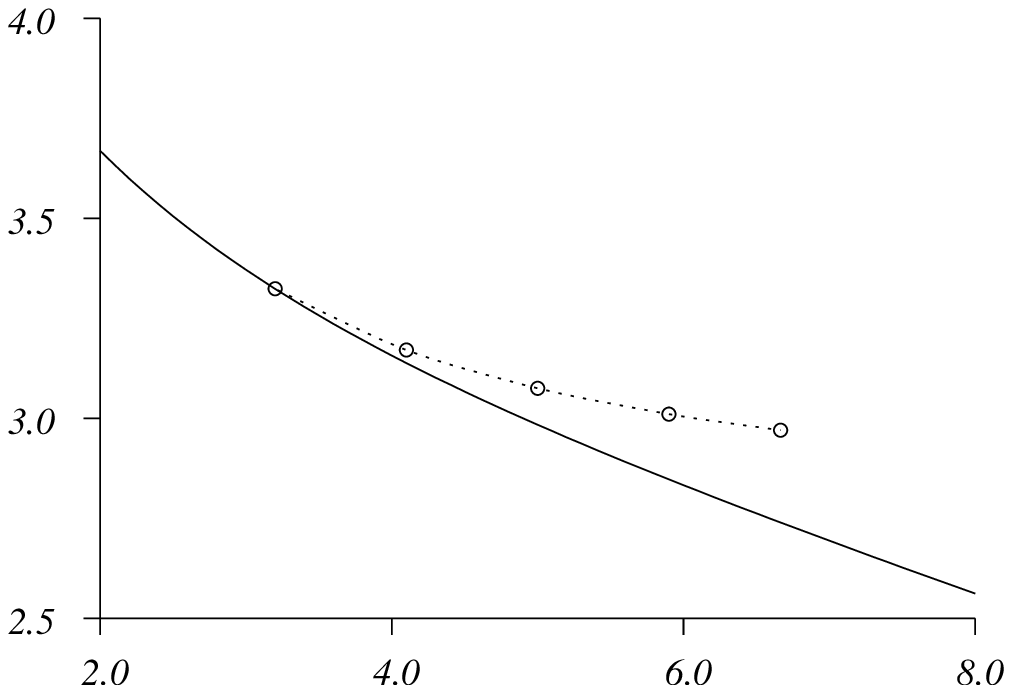}
\begin{picture}(0,0)
  \put(-1.0,2.5){$\alpha_W\,S$}
  \put(3.75,-0.5){$M_H / M_W$}
\end{picture}
\epsfysize=5cm
\epsfxsize=8cm
\epsfbox{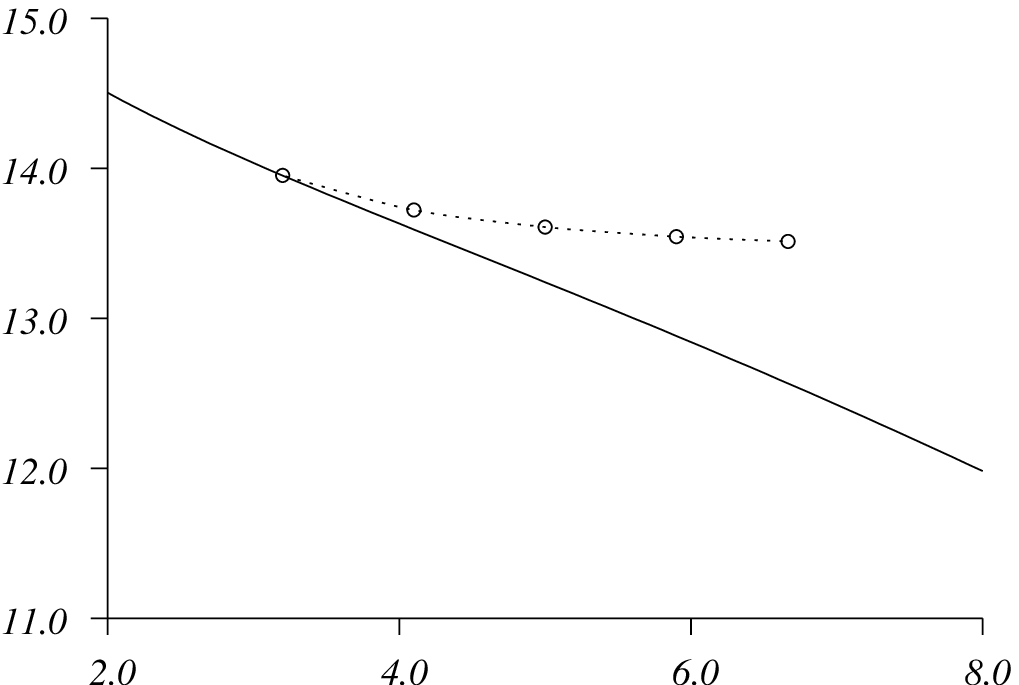}
\\[1.0cm]
\begin{picture}(0,0)
  \put(-2.0,2.5){$\alpha_W E / M_W$}
  \put(3.75,-0.5){$M_H / M_W$}
\end{picture}
\epsfysize=5cm
\epsfxsize=8cm
\epsfbox{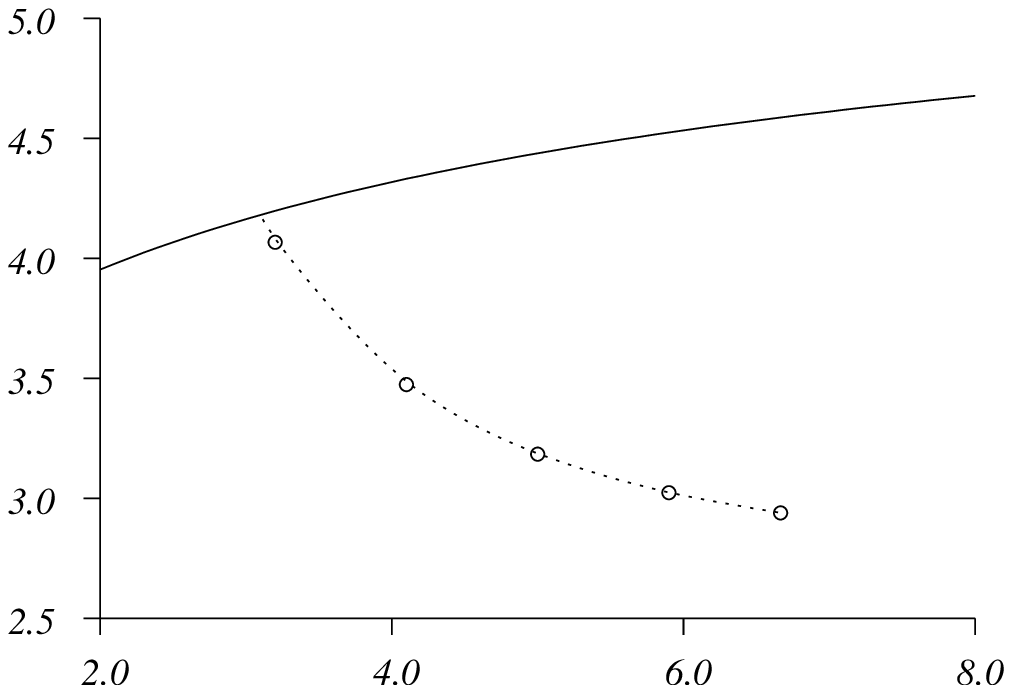}
\end{center}
\vspace{0.75cm}
\caption[Bifurcation Points vs $M_H$]{The calculated period
  $\beta$, action $S$, and turning point energy $E$ of the bounce
  bifurcation point $P$ are plotted as circles, as functions of the
  Higgs boson mass $M_H$.  The dotted lines are cubic spline fits to
  the calculated points.  The solid lines show the corresponding
  values of the sphaleron bifurcation point $Q$.
\label{pqvm-fig}}
\end{figure}

The period, action, and turning-point energy at the bifurcation points
$P$ and $Q$ are plotted in Fig.~\ref{pqvm-fig} as functions of the
Higgs boson mass $M_H$.  The bifurcation point $P$ was found by
increasing the period $\beta$ along each branch, using linear
extrapolation followed by Newton iteration, until this method failed
to converge no matter how small the step in $\beta$.  We confirmed
that the bifurcation was reached by comparing the last solution on
each branch.

\begin{figure}
\begin{center}
\setlength {\unitlength} {1cm}
\begin{picture}(0,0)
  \put(2.0,3.2){$|\phi|^2$}
  \put(3.8,2.0){$\mathrm{Re}(\chi^*\phi^2)$}
  \put(4.2,-0.5){$r$}
\end{picture}
\epsfysize=5.0cm
\epsfxsize=8.0cm
\epsfbox{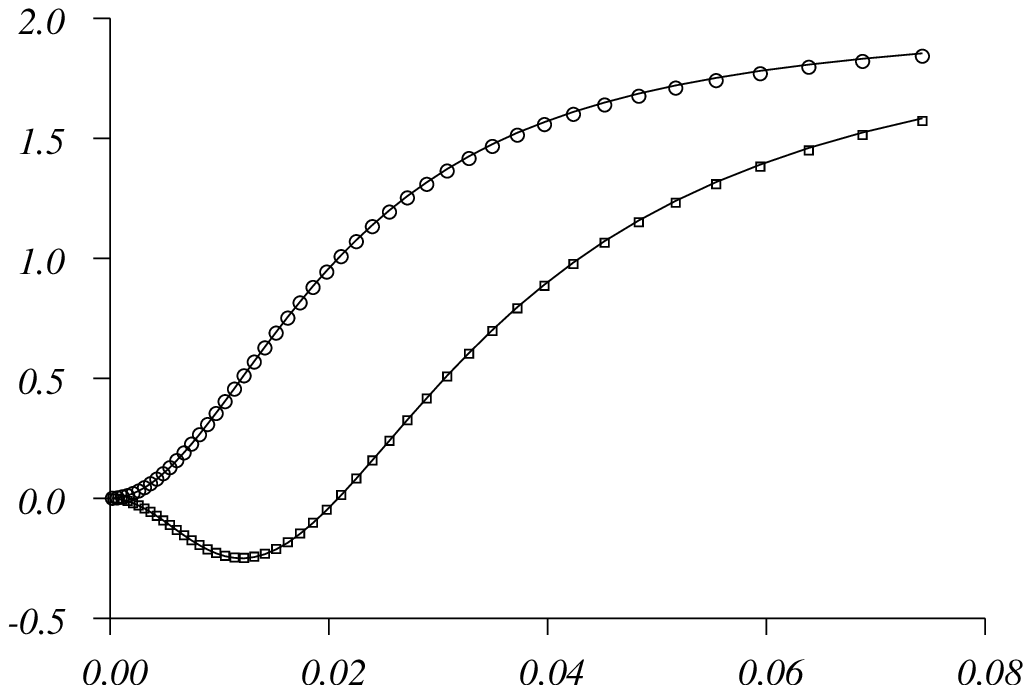}
\begin{picture}(0,0)
  \put(4.0,1.0){$|\chi|^2$}
  \put(3.8,4.2){$r^2f_{01}$}
  \put(4.2,-0.5){$r$}
\end{picture}
\epsfysize=5.0cm
\epsfxsize=8.0cm
\epsfbox{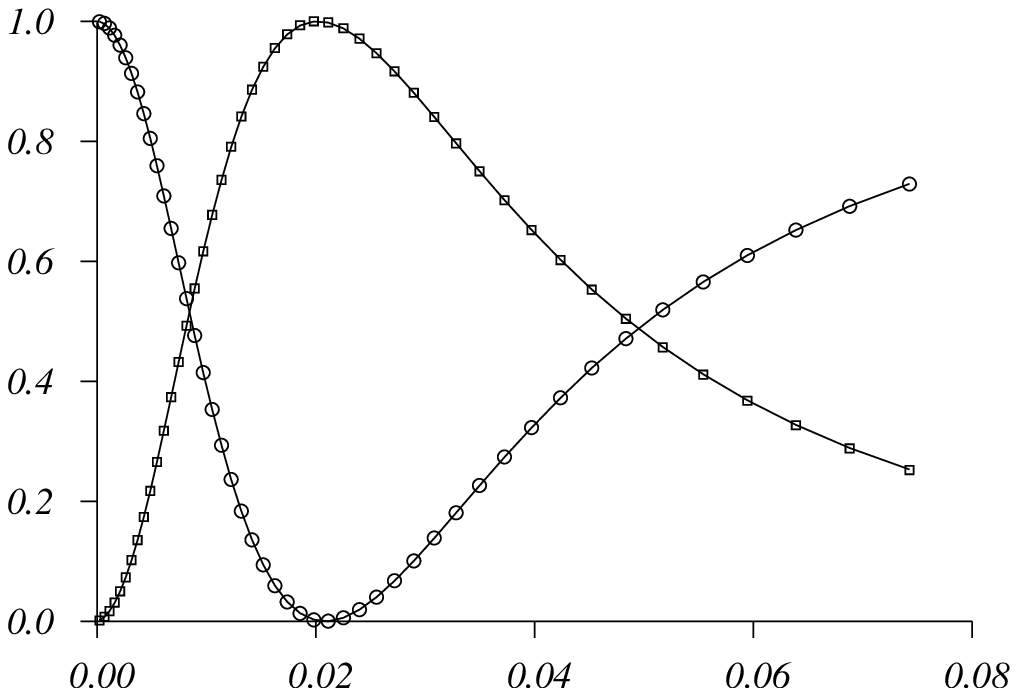}
\vspace{0.2cm}
\end{center}
\caption[Instanton Fields vs $r$]{The numerical bounce solutions approach
  instanton--anti-instanton fields, for small period ($\beta =
  0.75/M_W$).  The circles and squares show numerically calculated
  gauge-invariant quantities as functions of radius $r$, at a time
  slice midway between the turning points of the periodic solution, in
  units where $M_W=1$. The Higgs boson mass is $M_H = M_W$.  The left
  graph shows $|\phi|^2$ (circles) and $\mathrm{Re}(\chi^*\phi^2)$
  (squares); the right graph, $|\chi|^2$ (circles) and $r^2f_{01}$
  (squares).  The curves are from the analytic expressions
  (\ref{inst-2d}) for the instanton fields, with $\rho\approx
  0.0208/M_W$; the leading-order perturbative prediction is
  $\rho\approx 0.0202/M_W$.
\label{inr-fig}}
\end{figure}

\begin{figure}
\begin{center}
\setlength {\unitlength} {1cm}
\begin{picture}(0,0)
  \put(-1.0,3.0){$|\phi|^2$}
  \put(4.75,-0.5){$\tau - \beta/4$}
\end{picture}
\epsfysize=6.2cm
\epsfxsize=10.0cm
\epsfbox{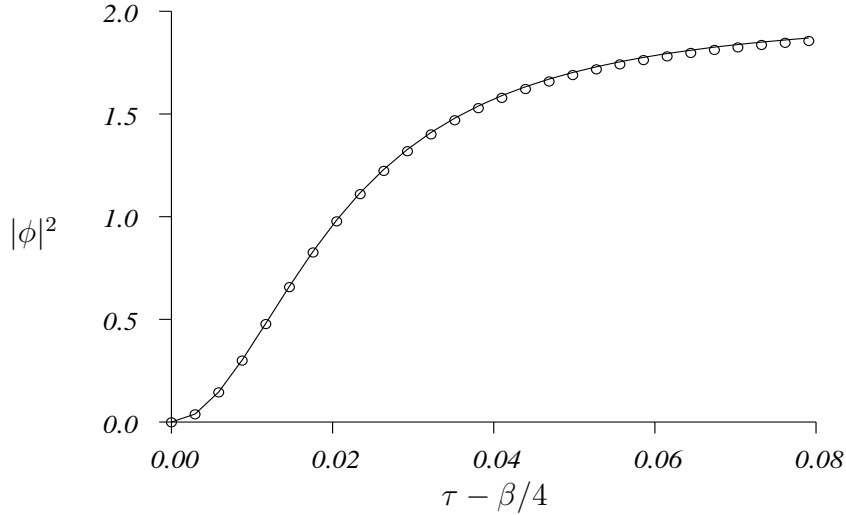}
\vspace{0.75cm}
\end{center}
\caption[Instanton Fields vs $\tau$]{The numerical bounce solutions
  approach instanton--anti-instanton fields, for small period
  ($\beta=0.75/M_W$).  The circles show the field $|\phi|^2$ as a
  function of imaginary time $\tau$, extrapolated to the spatial
  origin $r=0$, in units where $M_W=1$.  The Higgs boson mass is
  $M_H=M_W$.  The curve is from the analytic expression for the
  instanton field (\ref{inst-2d}), with the same value of $\rho$ as in
  Fig.~\ref{inr-fig}.
\label{int-fig}}
\end{figure}

At short periods $\beta \usebox{\lsim} 1/M_W$, the bounce solutions
approach small instanton--anti-instanton configurations.  The limit to
the accuracy of our computation in this region proved to be the number
of Fourier components in time we could afford.  With $N=64$
components, we could calculate with relative errors in the Euclidean
energy of no more than $\delta E / E \sim 10^{-3}$ down to periods
$\beta \approx 0.75/M_W$.  The calculated fields are compared to
analytic instanton fields [see Eqn.~(\ref{inst-2d}) below] in
Figs.~\ref{inr-fig} and~\ref{int-fig}.

For sufficiently short periods, one can find analytically the
asymptotic form of periodic Euclidean solutions of
$\mathrm{SU}(2)$-Higgs theory resembling chains of alternating
instantons and anti-instantons.  To leading order, the fields of an
instanton centered at the origin in singular gauge
are~\cite{tHooft_76}
\begin{mathletters}
\label{inst-eq}
\begin{eqnarray}
A_\mu(x) &=& \frac{2\rho^2}{x^2(\rho^2 + x^2)} \,
                \bar\eta^a_{\mu\nu}x_\nu (\tau^a / 2i)
                \> [1+{\mathcal O}(M_W x)],  \\
\Phi(x) &=& \frac{\sqrt {2} M_W \, x}{\sqrt{x^2+\rho^2}} \> \xi
                \> [1+{\mathcal O}(M_W x)+{\mathcal O}(M_H x)],
\end{eqnarray}
\end{mathletters}
where $\xi$~is an arbitrary complex unit doublet, and
$\bar\eta^a_{\mu\nu}$~denotes 't Hooft's
$\eta$~symbol~\cite{tHooft_76}.  Converting the instanton fields
(\ref{inst-eq}) into gauge-invariant two-dimensional quantities, we
find
\begin{mathletters}
\label{inst-2d}
\begin{eqnarray}
|\phi|^2 &=& \frac {2\,M_W^2} {1+\rho^2/x^2}\,,\kern 0.7in
f_{01} = \frac{4\rho^2}{[\rho^2+x^2]^2}\,,\\
|\chi|^2 &=&  1 - \frac{4r^2\rho^2}{[\rho^2+x^2]^2}\,,\kern 0.4in
\mathrm{Re}(\chi^*\phi^2) = |\phi|^2\,
\left( 1 - \frac{2r^2\rho^2}{x^2 [\rho^2 + x^2]} \right)\,.
\end{eqnarray}
\end{mathletters}
The agreement between the numerical results for $M_H=M_W$ and
$\beta=0.75/M_W$ and the asymptotic form (\ref{inst-2d}) shown in
Figs.~\ref{inr-fig} and~\ref{int-fig} convincingly demonstrates that
the numerical solutions have reached the instanton--anti-instanton
domain.

\section{Discussion}

At non-zero temperature, the leading exponential dependence of the
semi-classical barrier crossing rate is set by the action of the
periodic Euclidean
solution (with period $\beta$ equal to the inverse temperature) with
least action.
In electroweak theory, because of the anomaly in the baryon current,
this barrier crossing (or topological transition)
rate also gives the rate of baryon number non-conservation.
For sufficiently low temperature, this rate is determined by the limiting
action of zero-size instanton--anti-instanton configurations~%
\cite{tHooft_76,Frost&Yaffe_99,HabibMottola&Tinyakov_96}.  For higher
temperatures, the rate is set by the action of the static sphaleron
\cite{KuzminRubakov&Shaposhnikov_85,Arnold&McLerran_88}.  For Higgs
boson mass $M_H > 6.665\,M_W$, there is an intermediate range of
temperatures just below the temperature $1/\beta_0$ of the bifurcation
point $Q$, for which the periodic Euclidean solutions we calculated
numerically have a smaller action than the zero-size limit of
instanton--anti-instanton configurations.  In this range of
temperatures, the action of these periodic Euclidean solutions
determines the leading exponential behavior of the topological
transition rate.  As the temperature decreases, the action of the
time-dependent periodic solutions eventually exceeds the action of
zero-size instanton--anti-instantons.  At temperatures below this
crossing, the semi-classical transition rate is set by the zero-size
instanton--anti-instanton action.  For smaller Higgs boson mass $M_H <
6.665\,M_W$, non-singular time-dependent periodic solutions never
describe the dominant transition mechanism.  Instead, the leading
semi-classical transition rate undergoes an abrupt cross-over from
sphaleron-dominated transitions to zero-size instanton--anti-instanton
transitions, at the temperature where their actions cross.

Our numerical investigations were limited to real solutions of the
Euclidean field equations.  In addition to the real solutions we
calculated, one may easily show that there are also complex branches
of solutions which emerge from both bifurcation points $P$ and $Q$.
It is completely straightforward to extend the numerical techniques we
employed to find these complex branches of solutions.  This has been
done by Bonini \emph{et~al.}~\cite{Bonini&al_99}.  On the branch
emerging from the bifurcation $Q$ at the sphaleron, there is a single
``circle'' of complex solutions related
to one another by time translation symmetry.% 
\footnote{%
  To help understand the properties of these complex solutions, one
  may examine a simple algebraic model which reproduces the pattern of
  exact solutions, bifurcations, discrete symmetries, and negative and
  zero modes seen in the $\mathrm{SU}(2)$-Higgs model.  The toy model
  action is $S[x,y,z] = S_0 - z^2/2 + \Omega r^2/2 + \lambda r^4/4 -
  r^6/6$, where $r^2 \equiv x^2 {+} y^2$.  Rotations in the $x$-$y$
  plane correspond to Euclidean time translation.  The origin, $r = z
  =0$, represents the static sphaleron.  $z$ represents the amplitude
  of a deformation in the direction of the static negative mode of the
  sphaleron.  $x$ and $y$ represent amplitudes of the $\cos2 \pi
  t/\beta$ and $\sin 2\pi t/\beta$ components of the fundamental
  sphaleron oscillation, respectively.  Time reversal sends $y\to-y$
  leaving $x$ and $z$ unchanged.  Parity combined with a $\beta/2$
  time translation sends $z\to-z$ leaving $x$ and $y$ unchanged.
  Varying $\Omega$ is analogous to varying the period $\beta$ relative
  to $\beta_0$.  Varying $\lambda$ is analogous to varying the Higgs
  mass $M_H$ relative to $3.091 \, M_W$.  All the solutions we have
  discussed correspond to different extrema of $S$ for either real or
  complex $r^2$.  
} 
Both parity and time translation by $\beta/2$ act on this branch of
solutions as complex conjugation; therefore the action of these
complex solutions remains real, and decreases as one moves away from
the sphaleron.  These solutions are the analytic continuation to
Euclidean space of real Minkowski-space solutions whose energy is
greater than the sphaleron energy, and which are rolling toward, or
away from, the sphaleron.  Their precise initial conditions are
determined by the requirement that they have a periodic analytic
continuation to imaginary time.

For the other branch of complex solutions, emerging from the
bifurcation $P$, there are two different ``circles'' of solutions (not
related to one another by time translation) with complex action, and
both parity and complex conjugation map one circle of solutions onto
the other.  The physical relevance of these particular complex
solutions is not clear.  However, complex solutions satisfying
different boundary conditions are known to be relevant for
non-perturbative scattering~\cite{KhlebnikovRubakov&Tinyakov_91}, and
it would certainly be worthwhile to calculate those complex solutions
explicitly.

\section*{Acknowledgments}
The authors are grateful to S.~Habib, E.~Mottola, and R.~Singleton for
helpful discussions.  This work was supported in part by the U.S.
Department of Energy grant DE-FG03-96ER40956.

\bibliographystyle{prsty} \bibliography{perturb}

\begin{thebibliography}{10}

\bibitem{tHooft_76}
G. 't~Hooft, Phys.~Rev.~D {\bf 14},  3432  (1976).

\bibitem{Klinkhamer&Manton_84}
F.~R. Klinkhamer and N.~S. Manton, Phys.~Rev.~D {\bf 30},  2212  (1984).

\bibitem{KuzminRubakov&Shaposhnikov_85}
V.~A. Kuzmin, V.~A. Rubakov, and M.~E. Shaposhnikov, Phys.~Lett. {\bf 155B},
  36  (1985).

\bibitem{Arnold&McLerran_88}
P. Arnold and L. McLerran, Phys.~Rev.~D {\bf 37},  1020  (1988).

\bibitem{HabibMottola&Tinyakov_96}
S. Habib, E. Mottola, and P. Tinyakov, Phys.~Rev.~D {\bf 54},  7774  (1996).

\bibitem{Frost&Yaffe_99}
K.~L. Frost and L.~G. Yaffe, Phys.~Rev.~D {\bf 59},  065013  (1999).

\bibitem{Coleman_77}
S. Coleman, Phys.~Rev.~D {\bf 15},  2929  (1977).

\bibitem{KhlebnikovRubakov&Tinyakov_91}
S.~Y. Khlebnikov, V.~A. Rubakov, and P.~G. Tinyakov, Nucl.~Phys.~ {\bf B367},
  334  (1991).

\bibitem{Bonini&al_99}
G. Bonini {\it et~al.}, in Proc. of ``Strong and Electroweak Matter `98,''
  Copenhagen, December 1998.

\bibitem{Ratra&Yaffe_88}
B. Ratra and L.~G. Yaffe, Phys.~Lett.~B {\bf 205},  57  (1988).

\bibitem{Yaffe_89}
L.~G. Yaffe, Phys.~Rev.~D {\bf 40},  3463  (1989).

\bibitem{Kunz&Brihaye_89}
J. Kunz and Y. Brihaye, Phys.~Lett.~B {\bf 216},  353  (1989).

\end{thebibliography}

\end{document}